\documentclass[a4paper,english,aps,prl,twocolumn,superscriptaddress]{revtex4-1}
\usepackage[latin9]{inputenc}
\usepackage{amsbsy}
\usepackage{amssymb}
\usepackage{graphicx}
\usepackage{esint}

\makeatletter

\usepackage{color}
\usepackage{array}
\usepackage{textcomp}
\usepackage{bm}
\usepackage{multirow}
\usepackage{amssymb}
\usepackage{hyperref}
\hypersetup{colorlinks=true, citecolor=blue, urlcolor=blue, linkcolor=blue}
\usepackage{upgreek}
\usepackage{braket}

\makeatother

\usepackage{babel}
\begin{document}
\title{Impulsively Excited Gravitational Quantum States: Echoes and Time-resolved
Spectroscopy}
\author{I. Tutunnikov}
\affiliation{AMOS and Department of Chemical and Biological Physics, The Weizmann
Institute of Science, Rehovot 7610001, Israel}
\author{K. V. Rajitha}
\affiliation{AMOS and Department of Chemical and Biological Physics, The Weizmann
Institute of Science, Rehovot 7610001, Israel}
\author{A. Yu. Voronin}
\thanks{dr.a.voronin@gmail.com}
\affiliation{Lebedev Institute, 53 Leninsky pr., Moscow, Russia, Ru-119333}
\author{V. V. Nesvizhevsky}
\thanks{nesvizhevsky@ill.eu}
\affiliation{Institut Max von Laue-Paul Langevin (ILL), 71 avenue des Martyrs,
Grenoble, France, F-38042}
\author{I. Sh. Averbukh}
\thanks{ilya.averbukh@weizmann.ac.il}
\affiliation{AMOS and Department of Chemical and Biological Physics, The Weizmann
Institute of Science, Rehovot 7610001, Israel}
\begin{abstract}
We theoretically study an \textit{impulsively} excited quantum bouncer
(QB) - a particle bouncing off a surface in the presence of gravity.
A pair of time-delayed pulsed excitations is shown to induce a wave-packet
echo effect - a partial rephasing of the QB wave function appearing
at twice the delay between pulses. In addition, an appropriately chosen
observable {[}here, the population of the ground gravitational quantum
state (GQS){]} recorded as a function of the delay is shown to contain
the transition frequencies between the GQSs, their populations, and
partial phase information about the wave packet quantum amplitudes.
The wave-packet echo effect is a promising candidate method for precision
studies of GQSs of ultra-cold neutrons, atoms, and anti-atoms confined
in closed gravitational traps.
\end{abstract}
\maketitle
\emph{Introduction}---In last decades, massive quantum particle bouncing
off a surface under the influence of gravity turned from being an
issue of textbooks and pedagogical essays \citep{Haar1964,Langhoff1971,Gibbs1975,Banacloche1999}
into a subject of precision experiments on atom-optics gravitational
cavities \citep{Wallis1992,Andersen2002} and physics of ultra-cold
neutrons (UCNs) \citep{Nesvizhevskybook}. The observation of GQSs
\citep{Nesvizhevsky2002,Nesvizhevsky2003,Nesvizhevsky2005,Westphal2007}
and whispering gallery states \citep{Nesvizhevsky2008,Nesvizhevsky2010}
of neutrons ($n$) fueled a vast research in this area, which among
other goals, aims to the search for new fundamental short-range interactions
and physics beyond Standard Model, as well as verification of weak
equivalence principle in the quantum regime (see e.g. the introduction
of \citep{Nesvizhevsky2020}, and references therein).

Cold atoms and anti-atoms can also bounce on surfaces and form GQSs
\citep{Voronin2011} due to the quantum reflection from a rapidly
changing attractive van der Waals/Casimir-Polder surface potential
(see, e.g. \citep{Crepin_2019} and references therein). In contrast
to the extremely precise measurements of gravitational properties
of matter \citep{Adelberger1991,Darling1992,Huber2000}, the best
constraint \citep{Amole2013} for the gravitational mass (acceleration)
of antimatter does not allow even to define the sign of acceleration.
Several collaborations perform experiments at CERN \citep{Kellerbauer2008,Indelicato2014,Bertsche2018}
aiming to improve the accuracy. The GQS method seems to promise the
best accuracy for anti-hydrogen atoms $(\overline{\mathrm{H}})$ \citep{Crepin2019}.

Resonant spectroscopy of neutron GQSs was proposed in \citep{QGTrends2006},
measured using periodic excitation of QBs by mechanical vibrations
of the surface \citep{Abele2010,Jenke2011,Jenke2014,Cronenberg2018,Friedland2017},
and being implemented using a periodically changing magnetic field
gradient \citep{Antoniadis2014,Baessler2015}. Spatial distribution
of GQSs of $n$ was measured with micron resolution \citep{Ichikawa2015}.
For bouncing $\overline{\mathrm{H}}$ atoms, a resonant spectroscopy
\citep{Voronin2014,Crepin2017} and interferometry \citep{Voronin2016,Nesvizhevsky2019,Crepin2019}
approaches have been developed.

Here, we study the physics of \emph{impulsively} excited QBs, and
consider two example excitations: (i) by applying a pulsed magnetic
field gradient interacting with the QB's magnetic dipole moment, and
(ii) by a jolt caused by an impulsive shake of the surface. Short
laser pulses have been widely used for time-resolved molecular spectroscopy,
however the related aspects of the GQS spectroscopy are unexplored
yet. A spectacular effect in the dynamics of kick-excited nonlinear
systems is the echo phenomenon first discovered by E. Hahn in spin
systems \citep{Hahn1950,Hahn1953} (spin echo). Since then, various
types of echoes have been observed, including photon echoes \citep{Kurnit1964,Mukamel1995},
cyclotron echoes \citep{Hill1965}, plasma-wave echoes \citep{Gould1967},
neutron spin echo \citep{Mezei1972}, cold atom echoes in optical
traps \citep{Bulatov1998,Buchkremer2000,Herrera2012}, echoes in particle
accelerators \citep{Stupakov1992,Stupakov1993,Spentzouris1996,Stupakov2013,Sen2018},
and more recently---alignment and orientation echoes in molecular
gases \citep{Karras2015,Karras2016,Lin2016,Lin2020,Zhang2019,Ma2019,Rosenberg2018,Rosenberg2020,Hartmann2020}.
In these examples, echo appears in inhomogeneous ensembles of many
particles evolving at different frequencies. Echoes were also observed
in single quantum objects: in a single mode of quantized electromagnetic
field interacting with atoms passing through a cavity \citep{Morigi2002,Meunier2005},
and in single vibrationally excited molecules \citep{Qiang2020}.

In the first part of this Letter, we demonstrate, for the first time,
that highly nonlinear dynamics of quantum gravitational wave packets
favors observation of the echo in a \emph{single} QB. Then, we explore
a response of QB to a pair of time-delayed kicks, and analyze its
dependence on the delay between kicks. The population of ground GQS
as a function of the delay is shown to contain the transition frequencies
between the populated GQSs, as well as partial phase information about
the QB wave packet. This paves the way to a new kind of time-resolved
GQSs spectroscopy which has a number of advantages. It doesn't require
fine tuning of the excitation frequency to a specific resonance between
the GQSs, and eliminates some frequency shifts characteristic to the
resonant GQSs spectroscopy \citep{Baessler2015}.\\

\emph{Free quantum bouncer}---The vertical motion of the QB (along
$Z$ axis) is quantized and decoupled from the motion along the $X,Y$
axes. The eigenfunctions $\psi_{i}$ and energies $E_{i}$ of the
QB of mass $m$ are found from
\begin{equation}
H_{\mathrm{g}}\psi_{i}=-\frac{\hbar^{2}}{2m}\frac{\partial^{2}\psi_{i}}{\partial z^{2}}+mgz\psi_{i}=E_{i}\psi_{i},\label{eq:Shrodinger eq.}
\end{equation}
where $g$ is the gravitational acceleration, and $z$ is the vertical
position. Inertial and gravitational masses are taken as equal. The
perfect reflection off the surface is accounted for by the boundary
condition $\psi_{i}(z=0)=0$. The second boundary condition is $\psi_{i}(z\rightarrow\infty)=0$.
Position, time, and energy are measured in units of \citep{Banacloche1999}:
$z_{\mathrm{g}}=\left(\hbar^{2}/2m^{2}g\right)^{1/3}$, $t_{\mathrm{g}}=\hbar/E_{\mathrm{g}}$,
and $E_{\mathrm{g}}=mgz_{\mathrm{g}}$ (e.g. for neutron: $z_{\mathrm{g}}=5.87\;\upmu\mathrm{m}$,
$t_{\mathrm{g}}=1.094\;\mathrm{ms}$, $E_{\mathrm{g}}=0.60$ peV).
The solutions of Eq. (\ref{eq:Shrodinger eq.}) are shifted Airy functions
\citep{Banacloche1999}
\begin{equation}
\psi_{i}(z)=N_{i}\mathrm{Ai}(z-z_{i})=\frac{\mathrm{Ai}(z-z_{i})}{|\mathrm{Ai}^{\prime}(-z_{i})|},\label{eq:Dim. less. eig. funcs.}
\end{equation}
where $-z_{i}$ are the zeroes of $\mathrm{Ai}(z)$, and $N_{i}=|\mathrm{Ai}^{\prime}(-z_{i})|^{-1}$
are the normalization constants \citep{Albright1977}. The (positive)
energies are $E_{i}=z_{i}$ \citep{Banacloche1999}.\\

\begin{figure}
\centering{}\includegraphics[width=8.2cm]{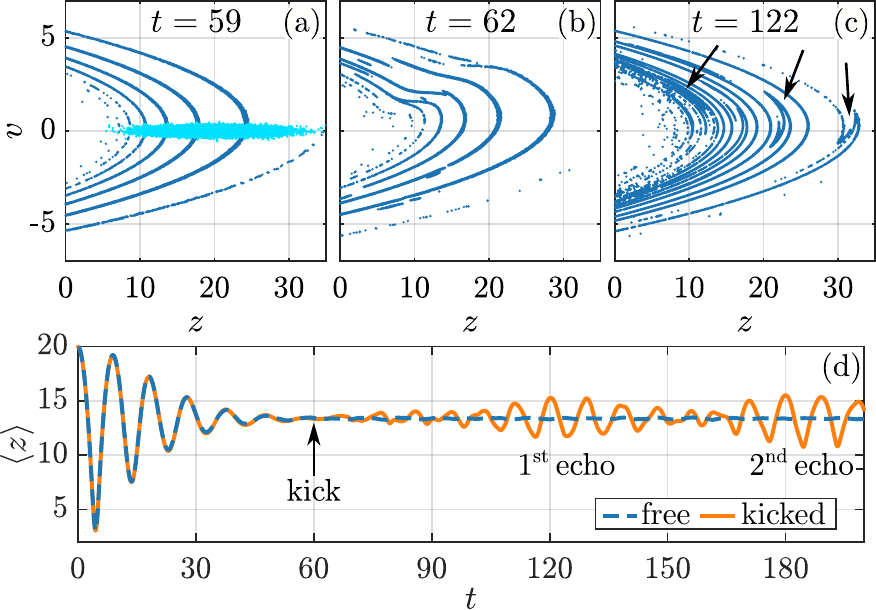}\caption{Phase space analysis. $N=2\times10^{4}$ particles bounce on a surface,
and kicked at $t=t_{\mathrm{k}}=60$. Kick parameters: $a_{\mathrm{k}}=0.5$
(e.g. for neutrons: $|\boldsymbol{\upmu}|=60.3$ neV/T, $\hat{\beta}\approx0.8$
T/m), $\sigma_{\mathrm{k}}=0.5$. Initial distribution {[}light blue,
(a){]} parameters: $\mu_{z}=20.0$, $\mu_{v}=0.0$, $\sigma_{z}=4$,
$\sigma_{v}=1/8$. (a) In blue - filamented phase space before the
kick. (b) Shortly after the kick. (c) Close to echo event, at $t\approx2t_{\mathrm{k}}$.
Arrows point at the tips (see the text). (d) Average position.\label{fig:Fig1}}
\end{figure}

\emph{Echo in a classical ensemble of gravitational bouncers}---It
is instructive to start from considering the dynamics of $N\gg1$
classical bouncing particles subject to a pair of delayed pulsed excitations
(``kicks''). The first kick initiates nonequilibrium dynamics in
the phase space. Here, for clarity of presentation, we model the resulting
phase space distribution by a displaced Gaussian with means $\mu_{z,v}$
and variances $\sigma_{z,v}$, for vertical position and velocity
{[}see the bright blue spot in Fig. \ref{fig:Fig1}(a){]}.

Due to the energy dependence of the bouncing frequency, the initial
smooth phase space distribution evolves into a spiral-like structure
{[}see the blue filaments in Fig. \ref{fig:Fig1}(a){]}. The number
of spiral turns increases with time, and they become thinner to conserve
the phase-space volume. Such ``filamentation'' is characteristic
of non-linear systems \citep{Lynden-Bell1967,Guignard1988,Stupakov2013}.
The spiral in the phase space exhibits itself via multiple sharp peaks
(``density waves'' \citep{Fisch2016}) in the QB's spatial distribution.

The filamented phase space serves as a basis for the echo formation
induced by the second kick applied at $t=t_{\mathrm{k}}$. Depending
on QB type and specific experimental implementation, various kicking
mechanisms can be utilized. As a first example here, we consider particles
with nonzero magnetic moment $\boldsymbol{\upmu}$, and kick them
using pulsed inhomogeneous magnetic field, $\mathbf{B}$. For simplicity,
we assume $\mathbf{B}$ has a uniform gradient near the surface \citep{Antoniadis2014,Baessler2015},
and fix $\boldsymbol{\upmu}$ along/against $\mathbf{B}$. Then, the
dimensionless interaction potential is $V_{\mathrm{B}}(z,t)=-s\beta(t)z$
$(s=\pm1)$, $\beta(t)=a_{\mathrm{k}}\exp[-(t-t_{\mathrm{k}})^{2}/\sigma_{\mathrm{k}}^{2}]$,
$a_{\mathrm{k}}=|\boldsymbol{\upmu}|\hat{\beta}/(mg)$, and $\hat{\beta}$
is the magnitude of the gradient. Figure \ref{fig:Fig1}(b) shows
the phase space distribution shortly after the kick, leading to particles
bunching, and formation of localized tips on each branch of the spiral.
The filamented structure provides a quasi-discrete set of oscillation
frequencies for the tips \citep{Lin2016,Lin2020}, which continue
evolving freely and, with time, get out of phase. However, due to
their quasi-discrete frequencies, the tips synchronize at twice the
delay, at $t\approx2t_{\mathrm{k}}$ {[}see Fig. \ref{fig:Fig1}(c){]},
resulting in the echo response \citep{Stupakov1992,Stupakov2013,Lin2016,Lin2020}.
Echo manifests in various physical observables. Here, we choose to
focus on the average position $\braket{z}(t)$ (also averaged over
$s=\pm1$). Figure \ref{fig:Fig1}(d) clearly shows the echo response
emerging at twice the kick delay, at $t\approx2t_{\mathrm{k}}$. Although
the tips fade with time, they synchronize quasi-periodically producing
higher order echoes \citep{Stupakov2013,Karras2016,Lin2016} visible
at $3t_{\mathrm{k}},4t_{\mathrm{k}}\dots$\\

\emph{Gravitational wave packet echo}---Initially, the QB is assumed
to be in a pure quantum state, e.g. a wave packet of GQSs. Pure GQS
has not been selected experimentally yet, due to tunneling of particles
through a gravitational barrier \citep{Nesvizhevsky2005}, but we
count on the major reduction of contamination of neighboring GQSs
in the future \citep{Escobar2014}. The QB may be set into motion
either by kicking it, or by dropping it on the surface from a step
\citep{Nesvizhevsky2004}/ion trap \citep{Crepin2019}. We start from
the latter, and model the initial state by a displaced Gaussian
\begin{equation}
\Psi(z,t=0)=\left(\frac{2}{\pi\sigma_{z}^{2}}\right)^{1/4}\exp\left[-\frac{(z-\mu_{z})^{2}}{\sigma_{z}^{2}}\right].\label{eq:Sec IV, magnetic initial state}
\end{equation}
This state is similar to the initial phase space distribution used
in the classical analysis. In the quantum case, the observable is
the expectation value, $\braket{z}(t)=\int_{0}^{\infty}\Psi^{*}(z,t)z\Psi(z,t)\mathrm{d}z$.
In principle, the echo effect can be observed in a variety of experimentally
accessible observables, e.g. a flux through the surface \citep{Nesvizhevsky2010a}.

\begin{figure}
\begin{centering}
\includegraphics[width=8.4cm]{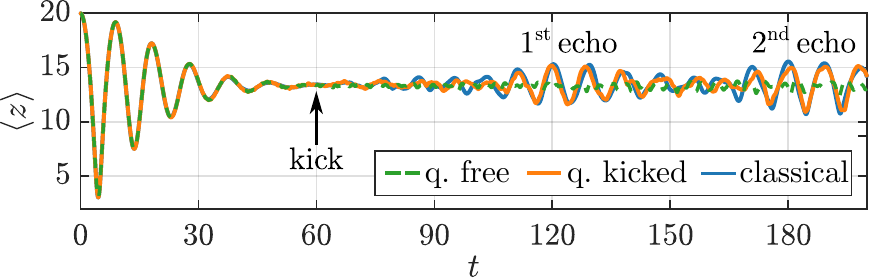}
\par\end{centering}
\caption{Echo induced by pulsed inhomogeneous magnetic field kick. The kick
is applied at $t=t_{\mathrm{k}}=60$. Initial state parameters: $\mu_{z}=20$,
$\sigma_{z}=8$ {[}see Eq. (\ref{eq:Sec IV, magnetic initial state}){]}.
Excitation parameters: $a_{\mathrm{k}}=\mu\hat{\beta}/mg=0.5$ (e.g.,
for neutrons: $\hat{\beta}\approx0.8$ T/m), and $\sigma_{\mathrm{k}}=0.5$.
The classical result {[}see Fig. \ref{fig:Fig1}(d){]} is added for
comparison. \label{fig:Fig2}}
\end{figure}

Figure \ref{fig:Fig2} shows that after several bounces, the wave
packet collapses {[}$\braket{z}(t)$ oscillations decay{]} because
of the differences in the transition frequencies of GQSs forming the
wave packet (a direct consequence of the anharmonicity of the potential).
The echo is induced by a kick applied after a delay $t_{\mathrm{k}}$.
Following the example considered classically, we assume that the QB
(an atom, anti-atom, or neutron) has spin $1/2$, and kick it by a
pulsed inhomogeneous magnetic field, $\mathbf{B}$. The Hamiltonian
is $H=H_{\mathrm{g}}-s\beta(t)z$, where $H_{\mathrm{g}}$ is defined
in Eq. (\ref{eq:Shrodinger eq.}), and $s=\pm1$ corresponds to the
spin states oriented along/against the field (see for details the
Supplemental Material).
The echo response is clearly visible at twice the kick delay, at $t\approx2t_{\mathrm{k}}$.
The result is the average of $\braket{z}(t)$ obtained for $s=\pm1$.

GQSs echo is conceptually different from classical echo emerging in
ensemble of many non-identical bouncers. The former can be observed
in single bouncers by repeating the experiment many times starting
from the same initial state. The interference pattern developing after
many measurements is a time-domain analogue of the spatial interference
fringes formed in the double slit experiment with single electrons
(the famous Feynman gedanken experiment, see \citep{Bach2013} and
references therein). Related echoes have been observed in single atoms
interacting with a single mode of cavity \citep{Morigi2002,Meunier2005},
and in single vibrationally excited molecules \citep{Qiang2020}.
The GQSs echo also differs from quantum revivals, which happen in
wave packets containing many states without additional kicks. The
periodicity of revivals depends only on the energy spectrum \citep{Eberly1980,Parker1986,Averbukh1989,Robinett2004},
while the echo period is controlled by the kick delay.\\

\begin{figure}
\begin{centering}
\includegraphics[width=8.2cm]{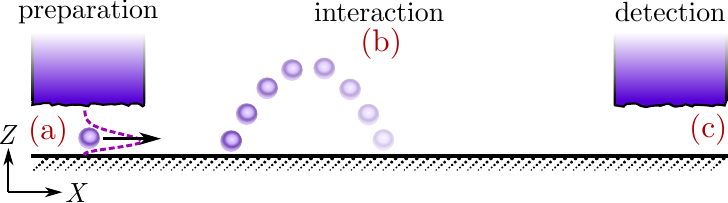}
\par\end{centering}
\caption{Schematic of a flow-through experimental setup. Bouncing (along $Z$
axis) particles propagate in $X$ direction. (a) First slit with a
rough top surface used for preparation. (b) Interaction region. (c)
Second slit used for detection. \label{fig:Fig3}}
\end{figure}

\emph{Time-resolved GQSs spectroscopy}---An appropriately chosen
observable measured \emph{as a function of the kick delay} contains
spectroscopic information about the QB. Here, for example, we choose
to follow the population of the ground GQS. The suggested measurement
can be realized in the typical flow-through configuration (see Fig.
\ref{fig:Fig3}) \citep{Nesvizhevsky2002,Jenke2011,Ichikawa2015,Roulier2015},
or using closed traps for QBs \citep{Roulier2015,Nesvizhevsky2020}.
The experiment includes three stages: preparation, interaction, and
detection. Initially, particles pass through a narrow slit {[}(a)
in Fig. \ref{fig:Fig3}{]}, whose top surface is rough leading to
the loss of highly excited particles. A sufficiently long and properly
sized slit allows preparing ground GQS, $\psi_{1}$ \citep{Luschikov1978,Nesvizhevsky2000}.
Then, the QB enters the interaction region {[}(b) in Fig. \ref{fig:Fig3}{]}
where it is subject to two kicks. In the detection stage, the particles
pass through the second slit {[}(c) in Fig. \ref{fig:Fig3}{]} allowing
only the population trapped in the ground state to reach the detector
(not shown). The delay, $\tau$ between the kicks is varied and the
population of the ground state is recorded as a function of $\tau$.

For impulsive (and identical) kicks, the Hamiltonian during the excitations
is $H\approx V(z)f(t)$. The wave function after the first kick is
given by $\Psi_{+}=\mathbf{P}\Psi_{-}$, where $\Psi_{-}$ is the
wave function before the kick, $\mathbf{P}=\exp[-i\alpha V(z)]$,
and $\alpha=\int_{-\infty}^{\infty}f(t)\mathrm{d}t$. For the initial
ground GQS, $\psi_{1}$, $\Psi_{+}=\sum_{i=1}^{\infty}P_{i1}\psi_{i}$,
where $P_{ij}$ is the matrix representation of $\mathbf{P}$ in the
basis of $\psi_{i}$s. After a delay $\tau$ (just before the second
excitation), the wave function is $\Psi_{-}(\tau)=\sum_{i=1}^{\infty}P_{i1}\psi_{i}e^{-iz_{i}\tau}$.
The \emph{delay-dependent} amplitude of the ground state after the
second kick is given by $c_{1}(\tau)=\sum_{i=1}^{\infty}P_{1i}^{2}\exp[-iz_{i}\tau]$,
while the population reads $|c_{1}|^{2}(\tau)=\sum_{i,j=1}^{\infty}(P_{1i}P_{1j}^{*})^{2}\exp\left[-i(z_{i}-z_{j})\tau\right].$
This signal oscillates at transition frequencies between the GQSs
populated by the first kick. The second kick affects the amplitudes
of the QGSs, but not their transition frequencies. In the limit of
weak kicks (keeping only terms with $i=1,j\geq1$ and $i\geq1,j=1$)
$|c_{1}|^{2}(\tau)$ reads
\begin{equation}
|c_{1}|^{2}(\tau)\approx\sum_{i=1}^{\infty}(P_{11}^{*}P_{1i})^{2}e^{-i(z_{i}-z_{1})\tau}+\mathrm{c.c.},\label{eq:GS-pop-afo-tau-(weak limit)}
\end{equation}
where ``c.c.'' stands for complex conjugate. The function in Eq.
(\ref{eq:GS-pop-afo-tau-(weak limit)}) contains the transition frequencies
between the excited states $\psi_{i}$ and the ground state $\psi_{1}$.
Notice that the signal contains \emph{phase information }allowing,
in principle, to retrieve the complex-valued wave function expansion
coefficients $P_{1i}$ (up to a $\pi$ phase). This is analogous to
the ``quantum holography'' procedure \citep{Leichtle1998,Averbukh1999,Weinacht1998,Weinacht1999}.
The access to phase information may open new possibilities for constraining
the parameters of extra interactions \citep{Abele2003,Baesler2007,Antoniadis2011,Brax2011}.

\begin{figure}
\begin{centering}
\includegraphics[width=8.2cm]{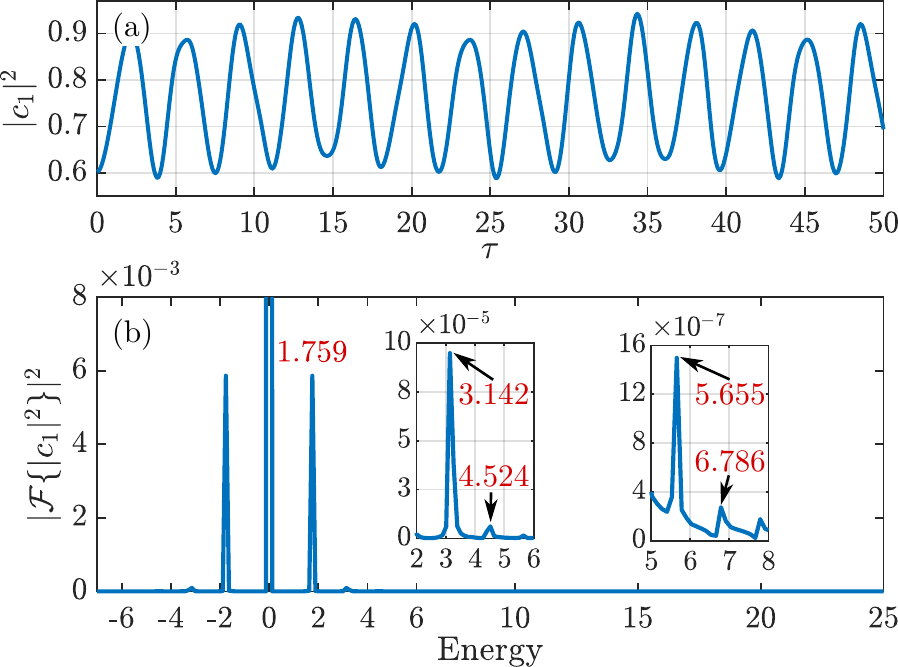}
\par\end{centering}
\caption{Time-resolved GQS spectroscopy: kicks by pulsed inhomogeneous magnetic
field\emph{.} (a) $|c_{1}|^{2}(\tau)$, kicks' parameters: $a_{\mathrm{k}1}=2$,
$a_{\mathrm{k}2}=1$ (e.g. for neutron: $|\boldsymbol{\upmu}|=60.3$
neV/T, $\hat{\beta}_{\mathrm{1}}\approx2.4$ T/m, $\hat{\beta}_{2}\approx1.2$
T/m), $\sigma_{\mathrm{k}1}=\sigma_{\mathrm{k}2}=0.2$. (b) Spectrum
of $|c_{1}|^{2}(\tau)$. Peaks correspond to energy differences $\mathcal{E}_{i1}$
($i=2,\dots,6$). Theoretical energy differences {[}see Eq. (\ref{eq:Dim. less. eig. funcs.}){]}:
$z_{21}=1.750,\,z_{31}=3.182,\,z_{41}=4.449,\,z_{51}=5.606,\,z_{61}=6.684$.
\label{fig:Fig4}}
\end{figure}

Figure \ref{fig:Fig4}(a) shows the numerically calculated $|c_{1}|^{2}(\tau)$
for the case of two delayed kicks by pulsed inhomogeneous magnetic
field. Here, the time dependence of the field is defined by $\beta(t)=a_{\mathrm{k}1}\exp[-t^{2}/\sigma_{\mathrm{k}1}^{2}]+a_{\mathrm{k}2}\exp[-(t-\tau)^{2}/\sigma_{\mathrm{k}2}^{2}]$.
The maximal delay is close to the typical time-of-flight through the
interaction region {[}(b) in Fig. \ref{fig:Fig3}{]} in experiments
with UCNs (see Ref. \citep{Antoniadis2014,Baessler2015} for details).
Figure \ref{fig:Fig4}(b) shows the spectrum of the signal in Fig.
\ref{fig:Fig4}(a), which contains mainly the energy differences $\mathcal{E}_{ij}=\mathcal{E}_{i}-\mathcal{E}_{j}$
between the low-lying excited states $\psi_{i}$ ($i=2,\dots,6$)
and the ground state $\psi_{1}$ {[}see Eq. (\ref{eq:GS-pop-afo-tau-(weak limit)}){]}.
The relative errors defined by $100\%\times(\mathcal{E}_{i1}-z_{i1})/z_{i1}$,
where $z_{i1}=z_{i}-z_{1}$ {[}see Eq. (\ref{eq:Dim. less. eig. funcs.}){]},
are $0.52\%,-1.27\%,\,1.69\%,\,0.87\%,\,1.52\%$ for $i=2,\dots,6$,
typical for flow-through experiments. The precision is determined
by the maximal delay, and may be increased in closed traps \citep{Nesvizhevsky2020}.\\

\emph{Kick by a jolt from the surface}---Both the wave packet echoes
and the QGSs spectroscopy approach discussed above are general and
do not depend on the specific type of kicks, as long as they are short.
Here we consider an additional kind of kicks caused by a sudden displacement
of the reflecting boundary. The corresponding Schr\"odinger equation
has a time-dependent boundary condition $\Psi[z=h(t)]=0$, where $h(t)$
is the surface height (see for details the Supplemental Material).
Such a model can, in principle, describe several experimental scenarios
in which the kicks are induced by shaking the surface as a whole,
or by existence of protrusions, grooves, or steps on the surface.
Such inhomogeneities appear as a time-dependent boundary in the reference
frame co-propagating transversally with the QB moving along the surface.
Here, $h(t)=a_{\mathrm{k}1}\exp[-t^{2}/\sigma_{\mathrm{k}1}^{2}]+a_{\mathrm{k}2}\exp[-(t-\tau)^{2}/\sigma_{\mathrm{k}2}^{2}]$,
where $a_{\mathrm{k}1},a_{\mathrm{k}2}$ are the amplitudes of the
kicks, and $\sigma_{\mathrm{k}1},\sigma_{\mathrm{k}2}$ define their
widths.

\begin{figure}
\centering{}\includegraphics[width=8.2cm]{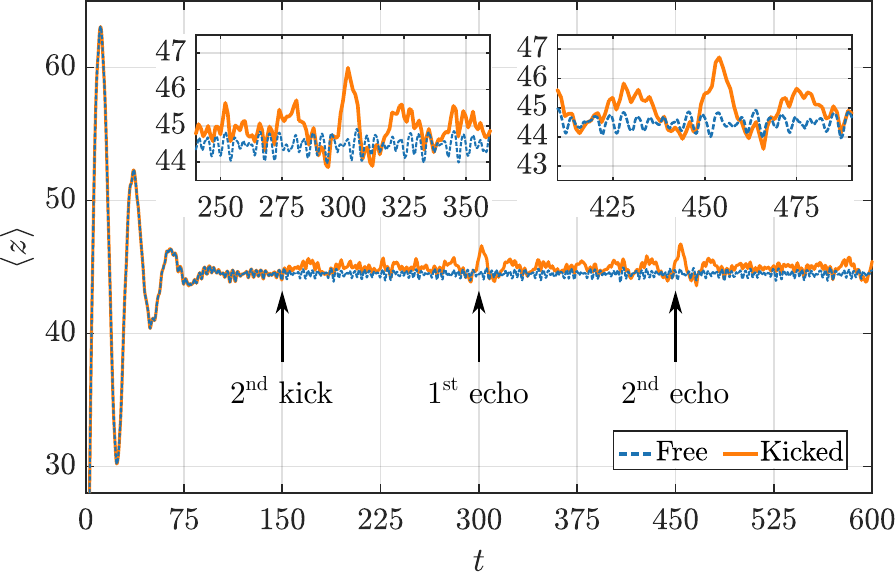}\caption{Echo induced by surface shake. The first kick is applied at $t=0$,
the delay of the second kick is $t_{\mathrm{k}}=150$. The echo emerges
at $t\approx2t_{\mathrm{k}},3t_{\mathrm{k}}$. Kicks' parameters:
$a_{\mathrm{k}1}=1.5$, $\sigma_{\mathrm{k}1}=1.0$, $a_{\mathrm{k}2}=0.10$,
$\sigma_{\mathrm{k}2}=0.16$. \label{fig:Fig5}}
\end{figure}

Figure \ref{fig:Fig5} shows the echo response of $\braket{z}(t)$.
Here, the QB is initially in the ground state $\psi_{1}$. A single
kick at $t=0$ excites a wave packet which collapses after several
oscillations (dashed blue). However, when a second kick is applied
at $t=t_{\mathrm{k}}$, echo responses emerge at $t\approx2t_{\mathrm{k}},3t_{\mathrm{k}}$
(solid orange).

\begin{figure}
\centering{}\includegraphics[width=8.2cm]{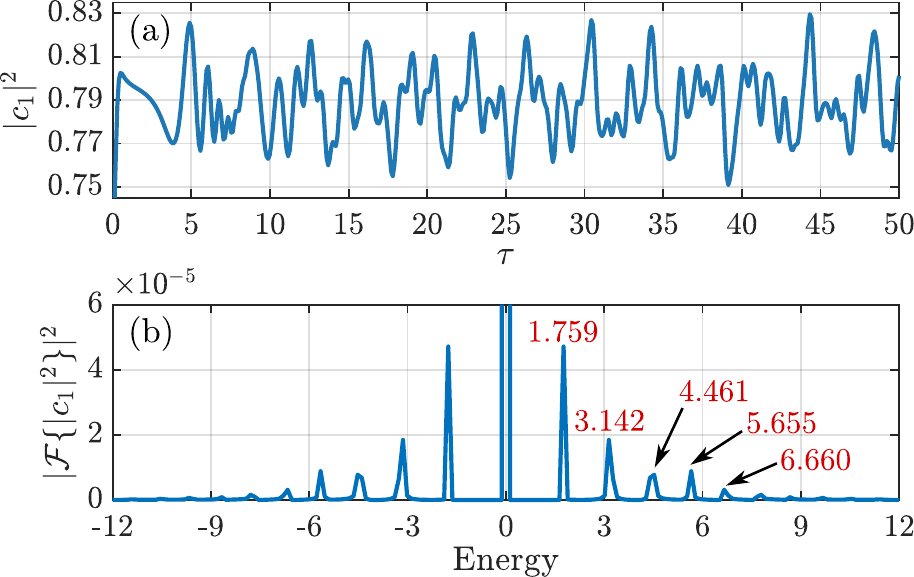}\caption{Time-resolved GQS spectroscopy: kicks by surface shake. (a) $|c_{1}|^{2}(\tau)$,
kicks' parameters: $a_{\mathrm{k}1}=0.6z_{\mathrm{g}}$, $a_{\mathrm{k}2}=0.1z_{\mathrm{g}}$,
and $\sigma_{\mathrm{k}1}=\sigma_{\mathrm{k}2}=0.2t_{\mathrm{g}}$.
(b) Spectrum of $|c_{1}|^{2}(\tau)$ shown in panel (a). Peaks correspond
to energy differences $\mathcal{E}_{i1}$ ($i=2,\dots,6$). Theoretical
differences {[}see Eq. (\ref{eq:Dim. less. eig. funcs.}){]}: $z_{21}=1.750,\,z_{31}=3.182,\,z_{41}=4.449,\,z_{51}=5.606,\,z_{61}=6.684$,
where $z_{i1}=z_{i}-z_{1}$. \label{fig:Fig6}}
\end{figure}

Figure \ref{fig:Fig6}(a) shows $|c_{1}|^{2}(\tau)$ in this case,
while the corresponding spectrum is shown in Fig. \ref{fig:Fig6}(b).
The maximal delay is close to the typical time-of-flight through the
interaction region in experiments with UCNs \citep{Nesvizhevsky2002,Ichikawa2015,Jenke2011,Roulier2015}.
The relative errors of the extracted energy differences are $0.52\%,\,-1.27\%,\,0.28\%,\,0.87\%,\,-0.37\%$
for $i=2,\dots,6$. In the limit of weak kicks ($a_{\mathrm{k1}},a_{\mathrm{k2}}\ll1$),
$|c_{1}|^{2}(\tau)$ can be obtained using time-dependent perturbation
theory {[}see Eq. (9) in the Supplemental Material{]}. In agreement
with Eq. (\ref{eq:GS-pop-afo-tau-(weak limit)}), the signal contains
the transition frequencies between the excited states $\psi_{i}$
and the ground state $\psi_{1}$, and the Fourier amplitudes are proportional
to the squared expansion coefficients of the wave packet after the
first excitation.\\

\emph{Conclusions.}---Echo effect in impulsively excited QBs is considered
and the formation mechanism is discussed using the auxiliary classical
model. Echoes may be used for probing decoherence effects originating
from interactions with the environment or other particles. The population
of the ground state recorded as a function of the delay is shown to
contain the transition frequencies between QGSs excited by the first
kick, populations, and partial phases information. The retrieved phases
may open opportunities for constraining the parameters of extra fundamental
interactions \citep{Abele2003,Baesler2007,Antoniadis2011,Brax2011}.
Various initial states, detection schemes, probe particles, and kicking
mechanisms can be envisioned for both inducing the echo effect and
QGSs spectroscopy. This method can be used by the current collaborations
working with GQSs of UCNs (Tokyo, qBounce, Los Alamos, GRANIT), with
$\overline{\mathrm{H}}$ (GBAR), with hydrogen atoms (GRASIAN), with
whispering gallery states of neutrons, atoms, and anti-atoms.\\

\begin{acknowledgments}
This work was supported in part by the Israel Science Foundation (Grant
No. 746/15). I.A. acknowledges support as the Patricia Elman Bildner
Professorial Chair. This research was made possible in part by the
historic generosity of the Harold Perlman Family.
\end{acknowledgments}

\end{document}